\documentclass[aps,pra,groupedaddress,showpacs,superscriptaddress,twocolumn]{revtex4}
\usepackage{graphicx}
\usepackage{amsmath,amssymb,amsfonts}
\usepackage{natbib}

\newcommand{\bd}{\begin{displaymath}}
\newcommand{\ed}{\end{displaymath}}
\newcommand{\be}{\begin{equation}}
\newcommand{\ee}{\end{equation}}
\newcommand{\bs}{\begin{subequations}}
\newcommand{\es}{\end{subequations}}
\newcommand{\ba}{\begin{eqnarray}}
\newcommand{\ea}{\end{eqnarray}}

\begin{document}

\title{Temperature crossover of decoherence rates\\
in chaotic and regular bath dynamics}

\author{A. S. Sanz}
\affiliation{Instituto de F\'{\i}sica Fundamental (IFF--CSIC),
Serrano 123, 28006 Madrid, Spain}

\author{Y. Elran}
\affiliation{The Weizmann Institute of Science, Rehovot, Israel}

\author{P. Brumer }
\affiliation{Chemical Physics Theory Group, Department of Chemistry, \\
and Center for Quantum Information and Quantum Control, \\
University of Toronto, Toronto, Ontario, Canada M5S 3H6.}


\begin{abstract}
The effect of chaotic bath dynamics on the decoherence of a quantum
system is examined for the vibrational degrees of freedom of a
diatomic molecule in a realistic, constant temperature collisional
bath. As an example, the specific case of I$_2$ in liquid xenon is
examined as a function of temperature, and the results compared with
an integrable xenon bath. A crossover in behavior is found: the
integrable bath induces more decoherence at low bath temperatures
than does the chaotic bath, whereas the opposite is the case at the
higher bath temperatures. These results, verifying a conjecture due
to Wilkie, shed light on the differing views of the effect of
chaotic dynamics on system decoherence.
\end{abstract}


\maketitle


\section{Introduction}
\label{sec1}

Decoherence, and the control of decoherence, is a central problem in
modern quantum physics. In particular, ``quantum technologies'',
such as quantum cryptography, quantum computing \cite{QC} and
quantum control \cite{SB} rely upon the maintenance of quantum
effects over significant periods of time. As such, decoherence
serves as a primary obstacle to progress in the experimental
implementation of a number of these quantum based scenarios. Of
particular interest is the nature and rates of decoherence in
systems in the condensed phase.

In this regard, it has been long argued
\cite{Zurek,Alicki,Zurek2,Tess,Pag,Man,Mil,You,Zhu,Zhu2,Cet,Harmon}
that traditional uncoupled oscillator or standard spin bath models
(e.g., the spin-boson \cite{Hu,Legg}, boson-boson
\cite{Feyn,Haake,Cald}, spin-spin \cite{Stamp} models) are
inadequate to describe dynamics in condensed phases since they lack
intra-environmental coupling. Such intra-environmental coupling can
display  different types of behavior, including chaotic dynamics,
and therefore its effect on decoherence can be significantly
different from traditional models.

For example, in such uncoupled bath cases, when the system perturbs
the bath, it cannot  relax  internally; energy must flow through the
subsystem in order for the bath to return to equilibrium. This
causes the system energy to increase initially even when relaxation
is expected \cite{Flem}. Second, the system becomes strongly
entangled with the bath as a result of this energy flow and hence
the system decoheres more strongly than it should. Finally, the
equilibrium state that the bath reaches may not be the expected
canonical state \cite{Can}.

Applications of uncoupled oscillators to model condensed phase
environments may be especially problematic. For example, anharmonic
corrections are known to be important in the study of phonons in Si
\cite{Wol} and are essential for the explanation of heat transport.
Both the vibrational dynamics of Si \cite{Miya} and its electronic
structure \cite{Mucc} are believed to be chaotic. Dynamics of a
colloidal particle in water have also been shown to be chaotic
\cite{Gasp}.

The effect of the structured environment on some issues in solid
state has been examined. For example, it has recently been shown,
for the central spin model
\cite{raedt,dukelsky,lai1,yang1,nie1,relano,erbe}, that the role of
structured environments on solid-state (ferromagnetic phase)
implementations is important. In particular,  the dynamical regime
of the bath has been observed \cite{dukelsky} to determine the
efficiency of the decoherence process. For example, in a
perturbative regime, decoherence is stronger in the integrable
limit; on the other hand, in the strong coupling regime  the chaotic
limit is more efficient. Also, the two-spin system decoherence has
been found to exhibit different behavior depending on the
characteristics of the coupling with the environment, as well as on
the internal dynamics and initial state of the environment.

Clearly, studies on more realistic models than the uncoupled
oscillator or spin models are needed for condensed phase
environments. Efforts to generalize the oscillator bath model in
such cases are very preliminary. When intra-bath coupling is added
to the boson or spin bath, general analytic solutions are
unavailable and exact solutions can only be obtained computationally
for very small baths. In the case of oscillator bath models, the
bath cannot consist of more than three or four oscillators. For spin
baths results have been reported for approximately 20 spins.

The most common question addressed in such studies is whether
intra-bath coupling increases or decreases the decoherence of the
embedded subsystem. Surprisingly, this apparently simple question
has generated considerable controversy. It was first conjectured by
Zurek that decoherence should be greater for chaotic baths
\cite{Zurek}. This was quantitatively verified in a study wherein a
single harmonic oscillator subsystem interacts with a bath
consisting of a single chaotic oscillator \cite{Zurek2}.
Unfortunately, the relevance of this result to the usual paradigm of
a small system interacting with a large environment is unclear.
Furthermore, Alicki has argued \cite{Alicki}, by contrast, that the
decoherence rate in the limit of pure decoherence (i.e. in the
absence of dissipation) will be greater for an integrable bath than
for a chaotic bath. This is because the energetically available bath
states in the integrable case can be highly degenerate, whereas only
one state is available in the chaotic bath at a given energy due to
level repulsion. Thus if a chaotic system-bath state is
microcanonical, the wavefunction of the system plus bath will be a
simple product state, since there is only  one energetically
available bath state with which  the system can couple. Accordingly,
there should be greater decoherence in the integrable case. This
viewpoint can also be supported with semiclassical arguments
\cite{Tess}. It can be shown that the square of the off-diagonal
matrix elements of the system-bath coupling operator scale as
$\hbar^{N-1}$ for a chaotic bath \cite{SemiC}. The off-diagonal
coupling matrix elements thus vanish in the thermodynamic limit. At
low temperatures the diagonal matrix elements change slowly with
energy \cite{Tess} and so the subsystem dynamics is shifted but not
strongly decohered. By contrast, selection rules for integrable
systems guarantee large off-diagonal matrix elements which cause
strong decoherence. These conclusions were verified numerically  for
a low temperature spin-bath \cite{Tess}.

Thus, there are two well-defined and apparently contradictory
positions on the issue of whether a chaotic bath may increase or
decrease decoherence.
Numerous low temperature spin-bath studies
\cite{Tess,Pag,Man,Mil,You,Zhu,Zhu2,Cet} support Alicki's
\cite{Alicki} predictions that decoherence should be greater for
integrable baths. However, there are spin-bath studies that draw the
opposite conclusion \cite{Harmon} and hence support Zurek's
conjecture that chaotic baths cause greater decoherence.

Based on these results Wilkie has speculated \cite{wilkieprivate}
that some sort of transition occurs with increasing temperature; a
chaotic bath could cause less decoherence at low temperature and
greater decoherence at high temperature. This conjecture would be
consistent with the spin-bath results
\cite{Tess,Pag,Man,Mil,You,Zhu,Zhu2,Cet,Harmon} and would not be in
direct conflict with the oscillator calculation \cite{Zurek2}.
However, the existence of such a transition is difficult to verify
in exact spin-bath or oscillator-bath calculations. High temperature
calculations for a bath of ten spins \cite{Tess} did not show such a
transition and calculations for larger spin-baths could not be
carried out at high temperature.

An alternative approach would be to explore the possibility of such
a transition using an approximation scheme.  This approach is the
focus of this paper. Recently it has been shown that quantum
decoherence can be accurately computed using classical dynamics
simulations based on the quantum Wigner function \cite{GB,elran1}.
In this Wigner approach, regions of phase space where the Wigner
function of the initial state takes negative values are Monte Carlo
sampled using the absolute value and the resulting classical
trajectories carry a negative sign as a weighting factor. The
resulting approximation can be very accurate, and in this paper
we employ this approach to examine the decoherence of a
superposition of vibrational states of I$_2$ in liquid Xe baths
comprising 512 atoms.

Our key observations are: (a) we observe less decoherence of
vibrational superposition states of I$_2$ at low temperatures for
liquid Xe than for its ideal gas counterpart obtained via
simulations without Xe-Xe interactions, but (b)  as the temperature
is increased, a transition is observed and liquid Xe becomes the
stronger source of decoherence. Thus, we show the existence of the
two regimes in a physically realistic model.  Note, as an immediate
application, that the decoherence of a vibrational superposition
state is a significant impediment to coherent control via pump-dump
scenarios \cite{SB}. Hence, understanding conditions responsible for
decoherence in such systems is important.  Indeed, this was the
original motivation for examining this particular system.

This paper is organized as follows. The model considered, as well as
the details of the numerical simulations, are discussed in
Sec.~\ref{sec2}. Section~\ref{sec3} reports the numerical results of
the simulations for two different initial states at three
temperatures and qualitative explanations for the observed behavior
are proposed. Finally, Sec.~\ref{sec4} contains a summary of
conclusions.


\section{Model}
\label{sec2}


\subsection{Hamiltonian}

Consider the decoherence rates for different superpositions of
vibrational states of  I$_2$ coupled to a bath of Xe atoms. The
subsystem of interest is the vibrational degree of freedom of the
diatomic, and the environment comprises  the translational degrees
of freedom of the  I$_2$ and of  the Xe atoms. The Hamiltonian
describing the full system \cite{egorov} can be written as a
standard system-plus-environment Hamiltonian, as follows:
\begin{equation}
 H = H_s + H_e + H_{se} ,
 \label{eq1}
\end{equation}
where
\begin{subequations}
 \begin{eqnarray}
 H_s & = & \frac{p^2}{2\mu_{\rm I_2}} + V (q) ,
  \label{eq2a} \\
 H_e & = & \frac{p_{\rm I_2}^2}{2m_{\rm I_2}}
  + \sum_i \frac{p_i^2}{2m_{\rm Xe}}
  + \sum_{i<j} \phi_{\rm Xe-Xe} (r_{ij}) ,
  \label{eq2b} \\
 H_{se} & = & \sum_i \phi_{\rm I_2-Xe} (r_{0i},q) .
  \label{eq2c}
 \end{eqnarray}
 \label{eq2}
\end{subequations}
Here, Eqs.~(\ref{eq2a}) and (\ref{eq2b}) describe the independent
evolution of the I$_2$ vibrational degree of freedom ($q$) and  the
bath dynamics, respectively, while Eq.~(\ref{eq2c}) accounts for
their interaction. The isolated I$_2$ is described by a Morse
oscillator,
\begin{equation}
 V = D \left[ 1 - e^{- \beta (q - q_0)} \right]^2 ,
 \label{eq22}
\end{equation}
with $D = 1.2547\times10^4$~cm$^{-1}$, $\beta = 1.8576$~\AA$^{-1}$,
$q_0 = 0$, and $\mu_{\rm I_2} = m_{\rm I_2}/4$, with $m_{\rm I_2}$
being the I$_2$ molecule mass and $\mu_{\rm I_2}$ being its reduced
mass. The degrees of freedom of the environment include the
translational degree of freedom of the I$_2$ (its center of mass,
with mass $m_{\rm I_2}$), as well as the collection of $N$ Xe atoms.
The interaction between Xe pairs is described by the Xe-Xe
interaction potential $\phi_{\rm Xe-Xe} (r_{ij})$, where $r_{ij}$ is
the distance between the $i$th and $j$th Xe atoms. This interaction
is modeled by a realistic pairwise Lennard-Jones potential:
\begin{equation}
 \phi_{\rm Xe-Xe} (r_{ij}) = 4 \epsilon_{\rm Xe-Xe}
  \left[ \left(\frac{\sigma_{\rm Xe-Xe}}{r_{ij}}\right)^{12}
   - \left(\frac{\sigma_{\rm Xe-Xe}}{r_{ij}}\right)^6 \right] ,
 \label{eq3}
\end{equation}
where $\epsilon_{\rm Xe-Xe} = 154.00$~cm$^{-1}$ is the well-depth of
the potential and $\sigma_{\rm Xe-Xe} = 3.930$~\AA\ is related to
the position of the minimum of the well [$V(r_{\rm min}) = -
\epsilon_{\rm Xe-Xe}$], $r_{\rm min} = 2^{1/6} \sigma_{\rm Xe-Xe}$.
Here, $\epsilon_{\rm Xe-Xe}$ gives an estimate of the intensity of
the interaction between two Xe atoms and $\sigma_{\rm Xe-Xe}$  is
the effective diameter of the Xe atoms. Note that under these
conditions and  at the densities and temperatures considered in this
work Xe is a liquid.

The coupling between the I$_2$ vibration and the bath is described
by the interaction Hamiltonian [see Eq.~(\ref{eq2c})]. If vibration
and translation were not coupled in the I$_2$, this term would just
account for the interaction between the I$_2$ with the Xe atoms and,
therefore, would look like Eq.~(\ref{eq3}), with $\epsilon_{\rm
Xe-Xe}$ and $\sigma_{\rm Xe-Xe}$ replaced by $\epsilon_{\rm I_2-Xe}$
and $\sigma_{\rm I_2-Xe}$, and $r_{ij}$ replacing $r_{0i}$ between
the I$_2$ and the $i$th Xe atom. In such a case, $\epsilon_{\rm
I_2-Xe}$ and $\sigma_{\rm I_2-Xe}$ can be taken as the average of
the corresponding Xe-Xe and I$_2$-I$_2$ interactions, i.e.,
$\epsilon = \sqrt{\epsilon_{\rm Xe-Xe} \epsilon_{\rm I_2-I_2}}$ and
$\sigma_{\rm I_2-Xe}^{(0)} = (\sigma_{\rm Xe-Xe} + \sigma_{\rm
I_2-I_2})/2$ (here, $\epsilon_{\rm I_2-I_2} = 382.27$~cm$^{-1}$ and
$\sigma_{\rm I_2-I_2} = 4.982$~\AA\ denote the well-depth and
position of the minimum for the corresponding I$_2$-I$_2$ pairwise
Lennard-Jones potential function). Here, however, since the diatomic
is ``breathing'' while it vibrates,  $\sigma$ plays the role of an
effective radius given by  $\sigma_{\rm I_2-Xe} = (\sigma_{\rm
Xe-Xe} + \sigma_{\rm I_2-I_2} + \alpha q)/2 = \sigma_{\rm
I_2-Xe}^{(0)} + \alpha q/2$ \cite{egorov}. Since diatoms expand and
contract in one direction, $\alpha \lesssim 1$. This model, termed
an effective breathing sphere,  models the interaction of I$_2$ with
the surrounding environment through the interaction potential:
\begin{equation}
 \phi_{\rm I_2-Xe} (r_{0i},q) = 4 \epsilon_{\rm I_2-Xe} \left[
  \left(\frac{\sigma_{\rm I_2-Xe}}{r_{0i}}\right)^{12}
   - \left(\frac{\sigma_{\rm I_2-Xe}}{r_{0i}}\right)^6 \right],
 \label{eq4}
\end{equation}
a potential that is known \cite{egorov} to be quantitatively
reliable for this system.

To obtain the analog of an uncoupled oscillator bath for comparison
purposes, we simply ignore the Xe-Xe interactions, setting the third
term in Eq.~(\ref{eq2b}) to zero. Collisional interactions between
the Xe atoms of the bath are thus removed, although collisions with
the I$_2$ are retained, a model  referred to below as the ``Xe ideal
gas''.


\subsection{Dynamics}

We consider initial states consisting of a thermally equilibrated Xe
bath within which is embedded an iodine molecule in a superposition
of vibrational states. Conceptually, such a superposition could be
prepared by laser excitation from the ground vibrational state,
where a multiphoton path can be utilized to overcome any selection
rule issues. The subsequent dynamics calculations are done by
sampling the Wigner distributions corresponding to the initial
superpositions of Morse eigenstates \cite{muendel} with classical
trajectories \cite{elran2}, followed by standard Molecular Dynamics
(MD) techniques \cite{frenkel} using the velocity-Verlet algorithm
\cite{allen} to propagate the trajectories. The result is the
time-evolving density $\rho(q,p,r_0,p_0,\{r_i,p_i\}_{i=1}^N,t)$ for
the full system + bath. A total number of $2\times10^6$ trajectories
was considered. In all reported simulations the total number of
particles (I$_2$ plus Xe atoms) is 512, which was found to be
adequate to converge the calculation of the purity, used as a
measure of the system coherence.  Finally, a density (I$_2$ in Xe)
$\rho^* = 3053$~g/cm$^3$ was  used in all the calculations to fix
the size of the MD cell.  Conversion factors and parameters used in
the simulation are provided in the Appendix.


\section{Computational Results and Discussion}
\label{sec3}

\begin{figure}
 \includegraphics[width=7cm]{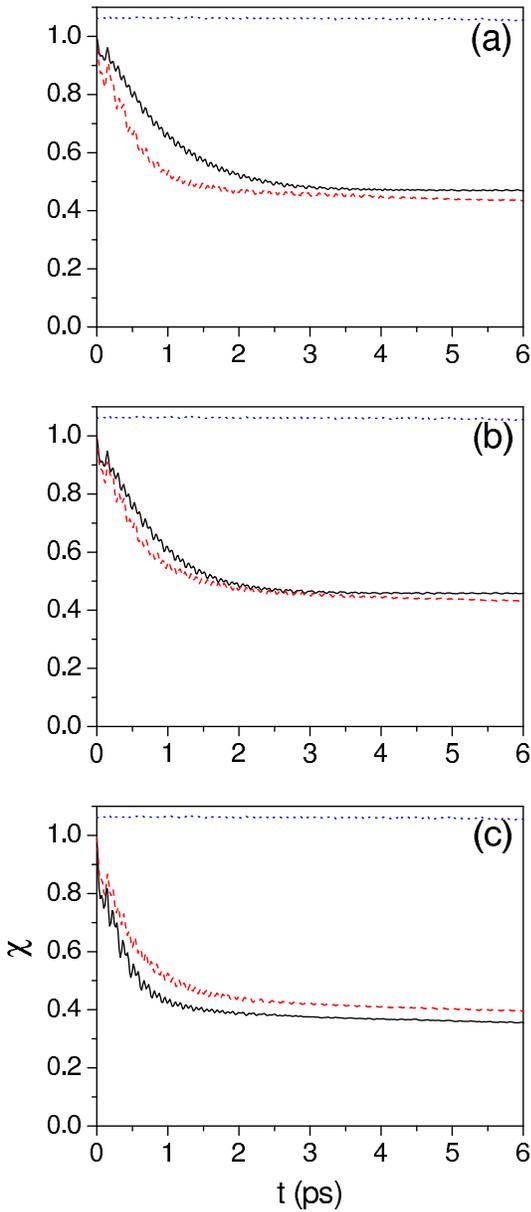}
 \caption{\label{fig1}
  Dependence on temperature of the purity as a function of time for the
  I$_2$ initially in a superposition of the ground and second-excited
  states: (a) $T = 177.36$~K, (b) $T = 221.7$~K and (c) $T = 554.25$~K.
  In all graphs, the black solid line indicates the Xe-Xe coupling is
  active, the red dashed line corresponds to the ``ideal gas'' bath (no
  intra-bath interactions), and the blue dotted line corresponds to the
  isolated I$_2$.}
\end{figure}

As a measure of decoherence, we compute the purity $\chi$ of the
I$_2$ dynamics \cite{xupei}, defined as
\be
 \chi = {\rm Tr}[\rho_s^2(t)] = \int \rho_s^2(q,p,t) dq dp ,
\ee
where
\be
 \rho_s(q,p,t) = \int \rho(q,p,r_0,p_0,\{r_i,p_i\}_{i=1}^N,t)
  dr_0 dp_0 \Pi_{i=1}^N dr_i dp_i
\ee
is the reduced density associated with the subsystem of interest,
which here is the I$_2$ vibrational degree of freedom. Since the
initial state (a vibrational superposition) is described by a
wavefunction, the purity is initially unity, but decays with time as
a consequence of the entanglement of the system and bath degrees of
freedom. A small amount of decay in $\chi$ is also observed as a
function of time for the isolated diatomic propagated in the absence
of the bath. This decay [see, e.g., Fig.~\ref{fig1}] is a measure of
the computational accuracy and is found to be very small over the
relevant 5~ps time scale.

Figure~\ref{fig1}(a) shows the purity as a function of time for the
two cases of Xe liquid and Xe ideal gas where the initial
vibrational degree of freedom of the I$_2$ is in an equal
superposition of the ground and second excited vibrational state. At
177.36 K both liquid Xe and ideal gas Xe are seen to cause
decoherence of I$_2$ on a picosecond time-scale. In this case it is
apparent that the liquid Xe bath causes substantially less
decoherence than does the ideal gas Xe bath.

\begin{figure}
 \includegraphics[width=7cm]{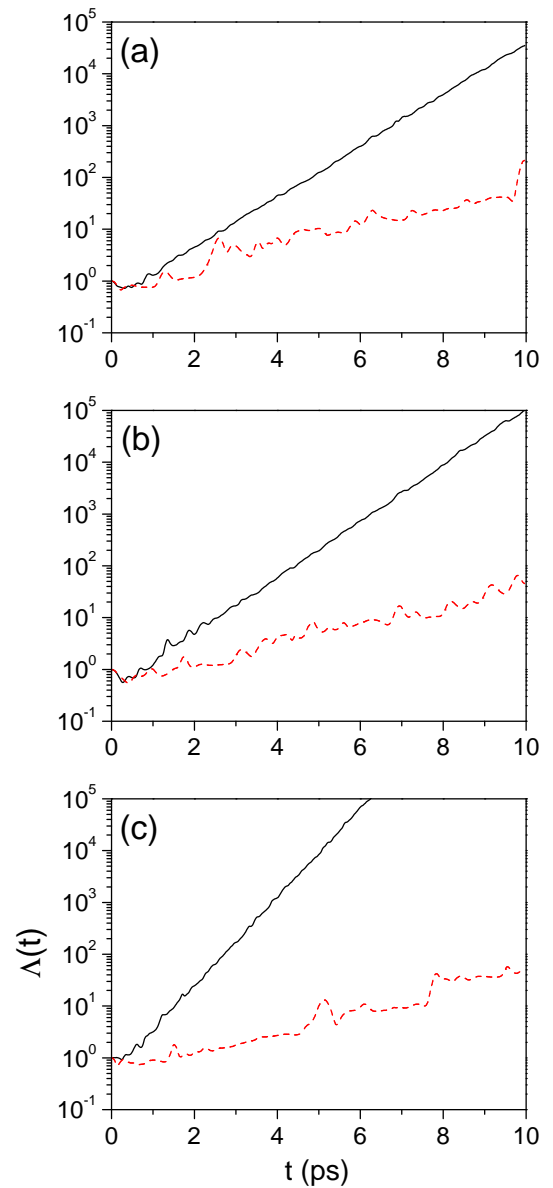}
 \caption{\label{fig2}
  Dependence on temperature of the Lyapunov exponent as a function of
  time for the I$_2$ initially in a superposition of the ground and
  second-excited states: (a) $T = 177.36$~K, (b) $T = 221.7$~K and
  (c) $T = 554.25$~K.
  In all graphs, the black solid line indicates the liquid Xe case
  and the red dashed line indicates the ideal gas Xe results.
  Note how $\Lambda(t)$ is similar in all three cases for the ideal
  gas Xe bath case.}
\end{figure}

To verify that the liquid Xe dynamics is indeed chaotic we
calculated the Lyapunov exponent, denoted by $\Lambda(t)$, as the
distance between two nearby Xe initial conditions \cite{PB}, which
is known to grow exponentially for a chaotic distribution dynamics
and sub-exponentially for integrable dynamics. Examination of the
Lyapunov exponent $\Lambda(t)$ as a function of time for the two
cases [see Fig.~\ref{fig2}(a)] clearly shows exponential growth for
liquid Xe and sub-exponential growth for ideal gas Xe. Hence, the
liquid Xe is chaotic, whereas the ``ideal gas'' case is not. The
results shown in Figure~\ref{fig1} support the Alicki conjecture
that bath chaos tends to suppress decoherence at low temperatures.

Similar results to those above are obtained at a temperature of
221.7~K. Figures~\ref{fig1}(b) and \ref{fig2}(b) show the purity and
Lyapunov exponents, respectively, plotted against time for the two
cases for the same 0-2 superposition. The liquid Xe shows the
expected exponentially increasing dependence of $\Lambda(t)$ with
time, showing that the bath is chaotic. Indeed, the slope of the
logarithm of $\Lambda(t)$ has increased, suggesting that the liquid
is even more chaotic than at the lower temperature. Again, the
liquid Xe case is found to induce less decoherence than does the
ideal gas Xe, but the difference between them is now less
pronounced.

The situation at high temperature, however, is quite different. As
seen in Fig.~\ref{fig1}(c), at 554.25 K  liquid Xe bath causes
greater decoherence than does the ideal gas Xe. The Lyapunov
exponent displayed in Fig.~\ref{fig2}(c) confirms that liquid Xe
bath is chaotic, with an even larger Lyapunov exponent [as manifest
in the slope of $\Lambda(t)$ versus $t$]. Thus, a transition has
occurred from a low temperature regime where the bath chaotic bath
induces less decoherence than does the non-collisional bath, to a
high temperature regime where the reverse is the case.

\begin{figure}
 \includegraphics[width=7cm]{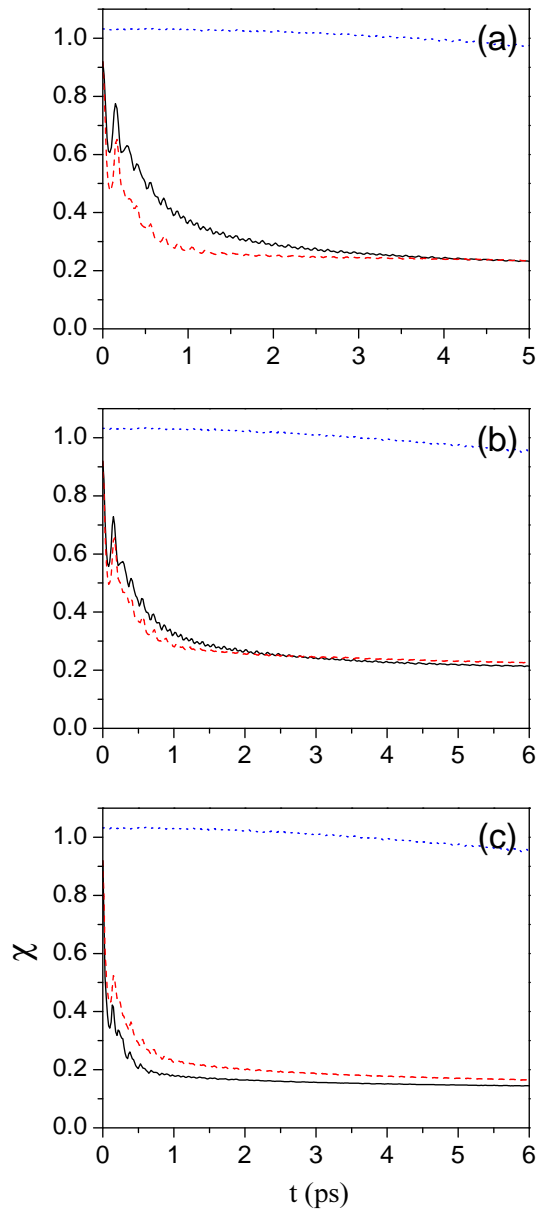}
 \caption{\label{fig3}
  Dependence on temperature of the purity as a function of time for the
  I$_2$ initially in a superposition of the fifth- and eighth-excited
  states: (a) $T = 177.36$~K, (b) $T = 221.7$~K and (c) $T = 554.25$~K.
  In all graphs, the black solid line indicates the Xe-Xe coupling is
  active, the red dashed line indicates the off coupling situation, and
  the blue dotted line indicates the isolated I$_2$.}
\end{figure}

It is evident from Fig.~\ref{fig3} that the decoherence dynamics for
the integrable ``ideal gas'' case is relatively unchanged as a
function of temperature. Hence, it is the chaotic case that goes
from weak to strong decoherence as the temperature increases, and is
responsible for the cross-over behavior in the decoherence of
regular versus chaotic baths.

To verify that this transition occurs for other initial conditions
we also show simulations for a second coherent vibrational
superposition state. Figures~\ref{fig3}(a)-(c) show the purity as a
function of time for three different temperatures (177.36, 221.7,
and 554.25~K) calculated for a coherent superposition of the fifth
and eighth vibrational states of the I$_2$. The decay of the purity
is faster for this second initial state than it is for the first.
Here the dynamics appears to occur on two time-scales, a fast
initial decay followed by a slower falloff. The fast initial decay
is common to both the liquid and ideal gas simulations. Otherwise
the plots are qualitatively similar to those for the first initial
state examined above. Significantly, the cross-over at high
temperatures is again apparent: Decoherence of the liquid is smaller
than that of the ideal gas at 177.36 and 221.7~K, but larger at
the highest temperature 554.25~K.


\section{Conclusions}
\label{sec4}

We have explicitly compared decoherence in the same system immersed
in a liquid Xe bath, a well known paradigmatic system. In doing so
we have confirmed that subsystems interacting with condensed phase
environments cannot be analyzed using approaches that neglect
intra-environmental coupling within the bath. Further, we
demonstrated that this coupling leads to a cross-over in the
decoherence dynamics as a function of temperature.

The transition described in the previous section has not been
previously observed computationally. It is clear that the integrable
bath case, in the liquid domain, is relatively insensitive to
temperature whereas the chaotic shows increasing decoherence with
increasing temperature. Since the chaotic bath case displays weaker
coherence than the integrable bath case at low temperature, the
increasing decoherence of the chaotic case with temperature results
in a cross-over of behavior. The low decoherence of the chaotic case
at low temperatures is in accord with the analysis in terms of
spectral properties of the bath. What is remarkable to note is that
the Wigner method is capable of properly displaying this behavior.

The results suggest a more detailed analysis of the origins of the
difference between the chaotic and integrable baths in different
density regions and the interrelationship between the rates of
decoherence and the Lyapunov exponent of the chaotic bath would be
of interest. Work on such systems is planned.


\begin{acknowledgments}

We thank Professor J. Wilkie for extensive discussions early in the
course of this work. Support from the Natural Sciences and
Engineering Research Council of Canada, and the Ministerio de
Econom{\'\i}a y Competitividad (Spain) under Projects FIS2010-22082 and
FIS2011-29596-C02-01, is acknowledged.
A.S.\ Sanz also thanks the Ministerio de Econom{\'\i}a y Competitividad
for a ``Ram\'on y Cajal'' Research Grant.

\end{acknowledgments}


\appendix
\section*{Computational aspects}
\label{appendix1}

The MD simulations were carried out using  well-known
standard procedures \cite{frenkel}. All the parameters
involved  were re-scaled
taking into account the parameters associated with the solvent
particles (here, the Xe atoms). That is:
\[ \begin{array}{ll}
 \mbox{interparticle distance:} & \qquad r^* = r/\sigma_{\rm Xe-Xe} , \\
 \mbox{time:} & \qquad t^* = \eta t , \\
 \mbox{frequency:} & \qquad \omega^* = \omega/\eta , \\
 \mbox{density:} & \qquad \rho^* = \sigma_{\rm Xe-Xe}^3 \rho , \\
 \mbox{temperature:} & \qquad T^* = k_B T/\epsilon_{\rm Xe-Xe} ,
\end{array} \]
with $\eta = \sqrt{\epsilon_{\rm Xe-Xe}/m_{\rm Xe}\sigma_{\rm
Xe-Xe}^2}^{1/2} \simeq 3.015 \times 10^{11}$~s$^{-1}$ (i.e., 1~MD
time unit is equivalent approximately to 3.32~ps) and where the
magnitudes with asterisk denote the re-scaled magnitudes. Thus, for
example, a density $\rho^* = 0.85$ will corresponds to $\rho =
3.053$~g/cm$^3$ and a temperature $T^* = 1.26$ to $T = 280$~K, while
Planck's constant will become $\hbar^* = \hbar/\sqrt{m_{\rm Xe-Xe}
\sigma_{\rm Xe-Xe}^2 \epsilon_{\rm Xe-Xe}} \simeq 0.010388$. This
scaling leads to a system of dimensionless equations of motion,
which are solved by means of the standard velocity-Verlet method in
the case of both the system and the environment. Within this scheme,
quantum features are taken into account initially in terms of the
classical Wigner method, i.e., by carrying out a Monte Carlo
sampling based on the Wigner distribution of the initial state of
the I$_2$ vibrational state.

The constants of solute and solvent are those previously obtained
for the Xe + I$_2$ system \cite{egorov}, i.e., $\epsilon_{\rm
I_2-I_2}/k_B = 550$~K, $\sigma_{\rm I_2-I_2} = 4.982$~\AA,
$\epsilon_{\rm Xe-Xe}/k_B = 221.7$~K and $\sigma_{\rm Xe-Xe} =
3.930$~\AA\ ($k_B = 1.3806505 \times 10^{-23}$~J/K being Boltzmann's
constant). The solute and solvent masses are, respectively, $m_0 =
4.22 \times 10^{-22}$~g and $m_s = 2.18 \times 10^{-22}$~g. Now,
taking into account the re-scaling, we will find $m_{\rm Xe}^* = 1$,
$\sigma_{\rm Xe-Xe}^* = 1$ and $\epsilon_{\rm Xe-Xe}^* = 1$ for the
Xe atoms, while $m_{\rm I_2}^* = 1.936$, $\sigma_{\rm I_2}^* =
1.268$ and $\epsilon_{\rm I_2}^* = 2.481$. Regarding the I$_2$
vibrational degree of freedom, we will find that $D^* = 81.4208$,
$\beta^* = 7.30037$ and $\mu_{\rm I_2}^* = 0.4839$. The vibrational
frequency corresponding to the I$_2$ is $\omega_{\rm I_2} =
\sqrt{2\beta^2 D/\mu_{\rm I_2}} = 4.0451 \times 10^{13}$~s$^{-1}$,
which in MD reduced units becomes $\omega_{\rm I_2}^* = 134.16$ in
reduced MD units (the period, in these MD units, is therefore $\tau
= 2\pi/\omega \simeq 0.0468$).


\end{document}